\documentclass[12pt]{article}
\usepackage{amssymb, amsmath, amsfonts,latexsym, verbatim, amsthm, mathrsfs,url}
\usepackage[usenames,dvipsnames]{color}

\title{Fermionic Wave Functions on Unordered Configurations}
\author{
Sheldon Goldstein\footnote{Departments of Mathematics, Physics and
     Philosophy, Rutgers University,  
     110 Frelinghuysen Road, Piscataway, NJ 08854-8019, USA.
     E-mail: oldstein@math.rutgers.edu},
James Taylor\footnote{Center for Talented Youth, Johns Hopkins University,
     McAuley Hall, Suite 400, 5801 Smith Ave, Baltimore, MD 21209, USA. E-mail:
     james.taylor@jhu.edu},\\
Roderich Tumulka\footnote{Department of Mathematics, 
     Rutgers University,  
     110 Frelinghuysen Road, Piscataway, NJ 08854-8019, USA.
     E-mail: tumulka@math.rutgers.edu},
\ and Nino Zangh\`\i\footnote{Dipartimento di Fisica dell'Universit\`a
     di Genova and INFN sezione di Genova, Via Dodecaneso 33, 16146
     Genova, Italy. E-mail: zanghi@ge.infn.it}
}
\date{March 11, 2014}

\newtheorem{defn}{Definition}
\newtheorem{prop}{Proposition}
\newenvironment{myproof}{\noindent\textit{Proof. }}{\hfill$\square$\bigskip}
\newenvironment{myproofof}[1]{\noindent\textit{Proof of #1. }}{\hfill$\square$\bigskip}
\newenvironment{ex}{\noindent\textit{Example. }}{\hfill$\square$\bigskip}
\newcounter{merken}

\newcommand{\inpr}[2]{\ensuremath{({#1}, {#2})}}
\newcommand{\dd}{\ensuremath{d}}
\newcommand{\rd}{\mathbb{R}^{\dd}}
\newcommand{\rdn}{\mathbb{R}^{\dd N}}
\newcommand{\nrd}{\ensuremath{{}^N\mspace{-1.0mu}\rd}}
\newcommand{\covspa}{\ensuremath{\widehat{\gencon}}}
\newcommand{\proj}{\ensuremath{\pi}}
\newcommand{\gencon}{\ensuremath{\mathcal{Q}}}
\newcommand{\deckt}{\ensuremath{\sigma}}
\newcommand{\covq}{\ensuremath{\hat{q}}}
\newcommand{\sspa}{\ensuremath{W}}
\newcommand{\herm}{Hermitian}
\newcommand{\nm}[1]{\ensuremath{{}^{N}\! {#1}}}
\newcommand{\id}{\ensuremath{\mathrm{Id}}}

\newcommand{\wh}[1]{\ensuremath{\widehat{#1}}}
\newcommand{\canq}{\ensuremath{\boldsymbol{x}}}

\newcommand{\rvarn}[1]{\ensuremath{\mathbb{R}^{#1}}}

\newcommand{\cid}{\ensuremath{\rvarn{\dd, N}_{\neq}}}

\DeclareMathOperator*{\lombdo}{\wedge}
\DeclareMathOperator*{\oplos}{\oplus}

\newcommand{\genset}{\ensuremath{T}}
\newcommand{\gena}{\ensuremath{a}}
\newcommand{\lilgenvec}{\ensuremath{w}}

\newcommand{\be}{\begin{equation}}
\newcommand{\ee}{\end{equation}}
\newcommand{\Hilbert}{\mathscr{H}} 
\newcommand{\Fourier}{\mathscr{F}} 
\newcommand{\Fock}{\mathscr{F}} 
\newcommand{\scp}[2]{\langle #1 | #2 \rangle} 
\newcommand{\Laplace}{\ensuremath{\Delta}} 
\newcommand{\pvm}{Q}
\newcommand{\domain}{\mathscr{D}}
\newcommand{\RRR}{\mathbb{R}} 
\newcommand{\CCC}{\mathbb{C}} 
\newcommand{\NNN}{\mathbb{N}} 

\newcommand{\vx}{\boldsymbol{x}}

\newcommand{\ve}{\boldsymbol{e}} 
\newcommand{\vA}{\boldsymbol{A}} 
\newcommand{\Q}{\gencon} 
\newcommand{\Euclideanspace}{\mathscr{E}} 

\newcommand{\holonomy}{\ensuremath{h}} 
\newcommand{\closedcurve}{\alpha}
\newcommand{\opencurve}{\beta}

\DeclareMathOperator{\Anti}{Anti} 
\DeclareMathOperator{\Sym}{Sym} 

\begin{document}
\maketitle

\begin{abstract}
Quantum mechanical wave functions of $N$ identical fermions are usually represented as anti-symmetric functions of \emph{ordered} configurations. Leinaas and Myrheim \cite{LM77} proposed how a fermionic wave function can be represented as a function of \emph{unordered} configurations, which is desirable as the ordering is artificial and unphysical. In this approach, the wave function is a cross-section of a particular Hermitian vector bundle over the configuration space, which we call the fermionic line bundle. 
Here, we provide a justification for Leinaas and Myrheim's proposal, that is, a justification for regarding cross-sections of the fermionic line bundle as equivalent to anti-symmetric functions of ordered configurations. In fact, we propose a general notion of equivalence of two quantum theories on the same configuration space; it is based on specifying a quantum theory as a triple $(\Hilbert,H,Q)$ (``quantum triple'') consisting of a Hilbert space $\Hilbert$, a Hamiltonian $H$, and a family of position operators (technically, a projection-valued measure on configuration space acting on $\Hilbert$).

\medskip

\noindent PACS. 03.65.Vf; 
03.65.Ta. 
Key words: bosons and fermions, symmetrization postulate, topological phases, Hermitian vector bundles, holonomy.
\end{abstract}

\newpage
\tableofcontents

\section{Introduction} \label{sec:intro}

For $N$ identical particles moving in $\RRR^d$, the appropriate configuration space is that of \emph{unordered} configurations,
\be\label{nrd}
\nrd := \{S \subset \RRR^d | \# S = N\}\,,
\ee
the set of all $N$-element subsets of $\RRR^d$. In quantum mechanics, one mostly uses the set of \emph{ordered} configurations,
\be
(\RRR^{d})^N\,,
\ee
the set of all $N$-tuples of elements of $\RRR^d$. For identical particles that space is inappropriate since identical particles are not ordered in nature; they are not labeled by numbers from 1 to $N$. However, it may seem that ordered configurations cannot be avoided for fermions, since for them the wave function is required by the \emph{symmetrization postulate} to be anti-symmetric,
\be
\psi(\canq_{\sigma (1)},\ldots, \canq_{\sigma (N)}) = 
(-1)^\sigma \, \psi(\canq_1,\ldots,\canq_N)\,,
\ee  
where $\sigma\in S_N$ is a permutation of $\{1,\ldots,N\}$, $S_N$ is the permutation group, $\canq_k\in\RRR^d$ for $k=1\ldots N$, and $(-1)^\sigma$ denotes the sign of $\sigma$. Leinaas and Myrheim \cite{LM77} proposed in 1977 a way how, for $d\geq 2$, ordered configurations can indeed be avoided, and $\psi$ can be regarded as a function on $\nrd$ as in \eqref{nrd}, by making it a cross-section of a suitable \herm\ vector bundle, the \emph{fermionic line bundle} $F(N,d)$. Such cross-sections are per se not in any way ``anti-symmetric,'' though they are in a one-to-one correspondence with the anti-symmetric functions on $(\RRR^d)^N$; this correspondence is canonical up to an overall phase factor. 

In this paper, we develop the approach of Leinaas and Myrheim further by making precise the sense in which cross-sections of the fermionic line bundle define \emph{the same quantum theory} as anti-symmetric wave functions on $(\RRR^d)^N$. In fact, we introduce a precise and generally applicable 
notion of what it means for two quantum theories to be equivalent, and go on to prove the equivalence between anti-symmetric wave functions on $(\RRR^d)^N$ and cross-sections of the fermionic line bundle over $\nrd$. 
Elsewhere \cite{GTTZ14b}, we discuss an analogous description of wave functions \emph{with spin}, for bosons or fermions, in terms of Hermitian vector bundles over the space $\nrd$ of unordered configurations. 

Let us describe our concept for when two quantum theories are equivalent. To this end, we define a \emph{quantum triple} to be a triple
\be
(\Hilbert,H, \pvm)
\ee
consisting of a Hilbert space $\Hilbert$, a self-adjoint operator $H$ in $\Hilbert$ called the Hamiltonian, and a PVM $\pvm$ (called the position observable) acting on $\Hilbert$ and defined on a measurable space $\gencon$ called the configuration space. We say that two quantum triples are \emph{equivalent},
\be\label{sim}
(\Hilbert,H,\pvm) \sim (\Hilbert',H',\pvm')\,,
\ee
if and only if $\pvm$ and $\pvm'$ are defined on the same configuration space $\gencon$ and there is a unitary isomorphism $U:\Hilbert\to\Hilbert'$ that maps $H$ into $H'$ and $\pvm$ into $\pvm'$, i.e., such that
\be\label{UHH'}
U(\domain(H))=\domain(H')\,, \quad
H'=UHU^{-1}\,, \quad
\text{and }\pvm'(A)=U\pvm(A)U^{-1}
\ee
for every measurable $A\subseteq \Q$, where $\domain(H)$ denotes the domain of $H$. 
As the name suggests, equivalence is an equivalence relation.

In Section~\ref{sec:equitriple} we will argue for the thesis that 
\newcounter{thesisnumber}
\setcounter{thesisnumber}{\theequation}
\begin{equation}\label{thesis}
\mbox{
\begin{minipage}{0.85\textwidth}
\textit{In order to completely specify the mathematical data defining a quantum theory, it is necessary and sufficient to specify an equivalence class of quantum triples.}
\end{minipage}}
\end{equation}
Put differently, a quantum triple provides sufficient mathematical data to uniquely define a quantum theory, and equivalent quantum triples are just different mathematical representations of the same quantum theory. 

The crucial mathematical observation in this paper, Proposition~\ref{prop:sim} below, is that the quantum triple over $\gencon=\nrd$ defined by the fermionic line bundle is equivalent to the quantum triple defined by anti-symmetric functions on $(\RRR^d)^N$. Our thesis \eqref{thesis} then yields that the two quantum triples characterize the same quantum theory, thereby justifying Leinaas and Myrheim's idea that wave functions that are cross-sections of the fermionic line bundle $F(N,d)$ provide just another mathematical description of the quantum mechanics of $N$ fermions in $\RRR^d$. Put differently, 
a fermionic wave function can indeed be represented by a function on $\nrd$, namely by a cross-section of $F(N,d)$.

\bigskip

The structure of this paper is as follows. Section 2 provides an overview of the definitions and results, while the other sections provide the proofs and other details. More specifically, in Section 2 we review the definition of the fermionic line bundle $F(N,d)$ for any number $N$ of particles and any number $d\geq 2$ of space dimensions and describe the link to anti-symmetric wave functions on $(\RRR^{d})^N$.
In Section~\ref{sec:fermionicbundle} we 
discuss various constructions and properties of $F(N,d)$. In Section~\ref{sec:anti} we 
prove that cross-sections of $F(N,d)$ are equivalent to fermionic wave functions. Section~\ref{sec:equitriple} is devoted to justifying our definition of equivalence of quantum theories given in \eqref{sim}. Section~\ref{sec:conclusions} offers conclusions.

\section{Overview}

We assume $d\geq 2$ throughout the paper, unless otherwise stated.

\subsection{Definition of the Fermionic Line Bundle}

Before we formulate the definition of the fermionic line bundle, we need to introduce some notation, terminology, and background. For a vector bundle $E$ over the base manifold $\Q$ with fiber spaces $E_q$, we also sometimes write $\cup_{q\in\Q} E_q$ (where the union is understood as a disjoint union).

\begin{defn}
    A \emph{\herm\ vector bundle}, or \emph{\herm\ bundle}, over a manifold
    $\Q$ is a finite-rank complex vector bundle $E$ over $\Q$
    with a connection and a positive-definite, Hermitian inner
    product $(\cdot,\cdot)_q$ on every $E_q$,
    the fiber of $E$ over $q\in\Q$, which is parallel.
\end{defn}

Such inner products are also called a \emph{fiber metric} (to be distinguished from a possible Riemannian metric on the tangent bundle).
That the inner product is parallel means that parallel transport (as defined by the connection) preserves inner products. Recall that the \textit{rank} of a vector bundle $E$ is the (complex) dimension of its fiber spaces $E_q$. A vector bundle of rank 1 is called a \textit{line bundle}. Note that since a \herm\ bundle consists of a vector bundle, a
connection, and a family of inner products, it can be nontrivial (i.e., not a product) even if the vector bundle is trivial:
namely, if the connection is nontrivial. In contrast, a \emph{trivial \herm\ bundle}, which we denote
$\Q \times \sspa$ for a Hermitian vector space (i.e., vector space with inner product) $\sspa$, consists of the trivial vector bundle (i.e., the Cartesian product of $\Q$ and $\sspa$) with corresponding inner products and the trivial connection, whose parallel transport $P_\opencurve$, in
general a unitary isomorphism from $E_q$ to $E_{q'}$ for $\opencurve$ a path
from $q$ to $q'$, is always the identity on $\sspa$. 
For any Hermitian bundle $E$ and any closed (piecewise smooth) curve $\closedcurve$ in $\Q$ beginning and ending at $q$, parallel transport defines a unitary endomorphism $\holonomy_\closedcurve: E_q \to E_q$, called the \emph{holonomy of $\closedcurve$}. A Hermitian bundle is called \emph{flat} iff the holonomy of every contractible closed curve is the identity; or, equivalently \cite[p.~92]{KN63}, iff the curvature is zero. 

\begin{defn}
An \emph{isomorphism $I$ of \herm\ bundles} $E$ and $E'$ over the same base manifold $\Q$ is a family of unitary isomorphisms $I_q:E_q\to E'_q$ for every $q\in\Q$ such that parallel transport along any curve in $\Q$ from $q$ to $r\in\Q$ will map $I_q$ into $I_r$.
\end{defn}

We note that, regarding $E$ as the set $\cup_q E_q$, we can also regard $I$ (i.e., the totality of the mappings $I_q$) as a mapping $I:E\to E'$. Also, since a linear mapping $E_q\to E'_q$ can be represented by an element of $E'_q\otimes E_q^*$, where $^*$ denotes the dual space, we can regard $I$ as a cross-section of $E'\otimes E^*$, the Hermitian bundle over $\Q$ whose fiber space at $q$ is $E'_q\otimes E^*_q$ with corresponding inner product and the connection inherited from $E$ and $E'$. It follows that this cross-section is parallel (i.e., has covariant derivative zero).

As we will explain below, for $d\geq 3$ there exist, up to isomorphism, two \emph{flat} Hermitian line bundles over $\nrd$. One is the trivial Hermitian line bundle $\nrd \times \CCC$, and the other is the fermionic line bundle $F(N,d)$; the latter is characterized by the property that the holonomy associated with a loop $\closedcurve$ in $\nrd$ starting and ending at $q$ is $\holonomy_\closedcurve = (-1)^{\sigma_\closedcurve}$, where $\sigma_\closedcurve$ is the permutation of $q$ defined by $\closedcurve$ and $(-1)^{\sigma_\closedcurve}$ its sign.  

To explain this, we begin with some facts about flat Hermitian bundles $E$ that will be relevant to us. The parallel transport $P_\opencurve$ along a path $\opencurve$ from $q$ to $q'$ depends only on the \emph{homotopy class} of $\opencurve$; i.e., if $\opencurve'$ is homotopic to $\opencurve$ then $P_{\opencurve'}=P_\opencurve$; that is because the concatenation $\opencurve^{-1}\opencurve'$ is null-homotopic (i.e., contractible) and by flatness has trivial holonomy. It follows that, for any fixed base point $q\in\Q$, the mapping $\gamma_{E,q}:[\closedcurve]\mapsto h_{\closedcurve}$ (where $\closedcurve$ is a closed curve based at $q$ and $[\closedcurve]$ its homotopy class) is a unitary representation of the fundamental group (first homotopy group) $\pi_1(\Q,q)$ with representation space $E_q$; it is called the \emph{holonomy representation}. Clearly, any Hermitian bundle $E'$ isomorphic to $E$ yields an \emph{equivalent}\footnote{We say that two unitary representations $\gamma_i:G\to U(W_i)$, $i=1,2$ of the group $G$, with $U(W_i)$ the unitary groups of the (possibly different) representation spaces $W_i$, are \emph{equivalent} iff there is a unitary isomorphism $U:W_1\to W_2$ such that $\gamma_2(g)=U\gamma_1(g)U^{-1}$ for all $g\in G$.} holonomy representation $\gamma_{E',q}$ of $\pi_1(\Q,q)$ on $E'_q$, viz., $\gamma_{E',q}([\closedcurve])= I_q\gamma_{E,q}([\closedcurve])I_q^{-1}$ for all $\closedcurve$. The converse is also true:

\newcounter{proprepresentation}
\setcounter{proprepresentation}{\theprop}
\begin{prop}\label{prop:representation}
For every connected differentiable manifold $\Q$, every $q\in\Q$, every $n\in\NNN$, and every $n$-dimensional unitary representation of the group $\pi_1(\Q,q)$, there exists a flat Hermitian bundle $E$ of rank $n$ over $\Q$ whose holonomy representation $\gamma_{E,q}$ is the given one. Any two such bundles are isomorphic; also, if two unitary representations of $\pi_1(\Q,q)$ are equivalent, then their associated flat Hermitian bundles are isomorphic. 
\end{prop}

This fact seems to be well known, but since we could not find an explicit formulation and proof in the literature, we include a proof in Appendix~\ref{app:representation}.

Let us return to the set $\nrd$.
It can be obtained from $(\RRR^d)^N$ (the space of ordered configurations) by first removing the collision configurations (with one or more particles at the same location) and then identifying configurations that are permutations of each other,
\be
\nrd \cong \cid/S_N \quad \text{with} \quad
\cid = \bigl\{ (\canq_1,\ldots, \canq_N) \in (\RRR^d)^N \big|
\canq_j \neq \canq_k \forall j\neq k \bigr\}\,.
\ee
Here, $\cong$ means a canonical correspondence. By means of this correspondence, since the permutation group $S_N$ acts properly discontinuously on $\cid$ by isometries, $\nrd$ is equipped with the structure of a Riemannian manifold. Since $\cid$ is connected, so is $\nrd$.

Next note that a closed curve $\closedcurve$ in $\nrd$ starting and ending at the configuration $q$ defines a permutation $\sigma_\closedcurve$ of $q$: Since every point in $\nrd$ corresponds to $N$ points in $\RRR^d$, the curve $\closedcurve$ defines $N$ curves in $\RRR^d$, each of which starts at a point $\vx$ belonging to $q$ and ends at a, possibly different, point $\sigma_\closedcurve(\vx)$ belonging to $q$. Since at each point belonging to $q$ one curve starts and one curve ends, the mapping $\sigma_\closedcurve: q\to q$ is bijective. 

If $d\geq 3$ then two closed curves in $\nrd$ based at $q$ are homotopic if and only if they define the same permutation. Thus, the fundamental group $\pi_1(\nrd,q)$ is isomorphic to the permutation group $S_N$. The 1-dimensional unitary representations $\gamma$ (also called \emph{characters}) of $S_N$ are well known: There exist two up to equivalence, the \emph{trivial character} $\gamma(\sigma)=1$ and the \emph{alternating character} $\gamma(\sigma) = (-1)^\sigma$ (the sign of $\sigma$). By Proposition~\ref{prop:representation}, there exist two flat line bundles over $\nrd$ up to isomorphism, $\nrd\times \CCC$ and the one corresponding to the alternating character. The latter is the definition of the \emph{fermionic line bundle} $F(N,d)$. The former (the trivial Hermitian line bundle) can be called the bosonic line bundle, as cross-sections of it can be regarded as bosonic wave functions; see \cite{GTTZ14b} for further discussion.

For $d=2$, the fundamental group of $\nrd$ is not isomorphic to $S_N$ but to the $N$-th braid group; this group possesses further characters besides the trivial and (the analog of) the alternating one (see Remark~\ref{rem:anyons} in Section~\ref{sec:rem} for further discussion). Anyway, we define the fermionic line bundle $F(N,2)$ to be the Hermitian line bundle corresponding to the alternating character. That is:

\begin{defn}
For any $N\in\NNN$ and $d\geq 2$, the \emph{fermionic line bundle $F(N,d)$} is the (unique up to isomorphism) Hermitian bundle over $\nrd$ with holonomies $h_{\closedcurve}=(-1)^{\sigma_\closedcurve}$.
\end{defn}

In Section~\ref{sec:fermionicbundle} we describe several ways of constructing, or realizing, $F(N,d)$.
The following remarks about $F(N,d)$ will be proved in Section~\ref{sec:rem}.
The fermionic line bundle for $d \geq 3$ possesses not only a
nontrivial connection, it is also nontrivial as a vector bundle
(disregarding connection and Hermitian inner products). As a consequence,
every smooth cross-section has nodes, i.e., it vanishes somewhere. (That is
because any nowhere-vanishing smooth cross-section of a line bundle would define a
global trivialization of the bundle.) A further consequence is that $\nrd$ is not
orientable for odd $d$.

\subsection{Correspondence to Anti-Symmetric Wave Functions}

We now explain how cross-sections of $F(N,d)$ correspond to anti-symmetric functions on $(\RRR^d)^N$; proofs and further details will be given in Section~\ref{sec:anti}. Let us fix $N$ and $d$ and simply write $F$ for $F(N,d)$. Let $\proj:\cid \to \nrd$ be the projection mapping from ordered to unordered configurations,
\be\label{projdef}
\proj(\canq_1,\ldots,\canq_N) = \{\canq_1,\ldots,\canq_N\}\,.
\ee 
Note that $\proj$ is a covering map, with the covering fiber $\proj^{-1}(q)$ over an unordered configuration $q$ given by the set of all possible orderings of $q$. As a consequence, for any Hermitian bundle $E$ over $\nrd$ there is a corresponding Hermitian bundle $\wh{E}=\proj_*(E)$ over $\cid$, the \emph{pull-back} of $E$ through $\proj$, or \emph{lift} of $E$. 

\newcounter{prophatFtrivial}
\setcounter{prophatFtrivial}{\theprop}
\begin{prop}\label{prop:hatFtrivial}
For any $N\in\NNN$ and $d\geq 2$, the lift $\wh{F} = \proj_* F$ of the fermionic line bundle to $\cid$ is a trivial Hermitian bundle.
\end{prop}

Choose a global trivialization of $\wh{F}$ (together with its Hermitian metric!); that is, choose an isomorphism of Hermitian bundles $\hat{I}:\wh{F} \to \cid \times \CCC$. Any two choices of $\hat{I}$ differ at most by a global phase factor, $\hat{I}_2=e^{i\theta}\hat{I}_1$ for some $\theta\in\RRR$. Using this trivialization, we can regard cross-sections of $\wh{F}$ as complex-valued functions on $\cid$. For any cross-section $\psi$ of $F$, its pull-back $\proj_* \psi$ is a cross-section of $\wh{F}$, and by applying $\hat{I}$ (and a normalizing factor $N!^{-1/2}$) we obtain a complex-valued function $U\psi$ on $\cid$: namely, for $\covq\in\cid$, let $\hat{I}_{\covq}: \wh{F}_{\covq} \to \{\covq\}\times\CCC$ be the restriction of $\hat{I}$ to the fiber space over $\covq$, for which we simply write $\hat{I}_{\covq}: \wh{F}_{\covq} \to\CCC$ as it is a linear mapping. In this notation,
\be\label{whpsihatI}
\bigl(U\psi\bigr)(\canq_1,\ldots,\canq_N) = 
\frac{1}{\sqrt{N!}}\hat{I}_{(\canq_1\ldots\canq_N)} \psi(\{\canq_1,\ldots,\canq_N\})\,.
\ee
(The factor $1/\sqrt{N!}$ serves for making $U$ unitary, see Proposition~\ref{prop:smooth} below, by compensating for the $N!$-fold overcounting of configurations when integrating $|U\psi|^2$ over $\RRR^{Nd}$.)

\newcounter{propanti}
\setcounter{propanti}{\theprop}
\begin{prop}\label{prop:anti}
For any $N\in\NNN$ and $d\geq 2$, any global trivialization $\hat{I}:\wh{F} \to \cid \times \CCC$, and any cross-section $\psi$ of $F$, $U\psi$ as given by \eqref{whpsihatI} is an anti-symmetric function on $\cid$. 
\end{prop}

It is useful to formulate the correspondence between cross-sections of $F$ and anti-symmetric functions in terms of Hilbert spaces. We will simply write $\RRR^{Nd}$ in the following for $(\RRR^d)^N$. Let ``$\Anti$'' denote the anti-symmetrization operator, defined for 
any function $\wh{\psi}:\RRR^{Nd}\to\CCC$ by
\be\label{Antidef}
\Anti \wh{\psi}(\canq_1,\ldots,\canq_N) = 
\frac{1}{N!}\sum_{\sigma \in S_N} (-1)^\sigma
\wh\psi(\canq_{\sigma(1)},\ldots, \canq_{\sigma (N)})\,. 
\ee
We write $\Anti C^\infty(\RRR^{Nd})$ for the image of $C^\infty(\RRR^{Nd})$ under $\Anti$, i.e., for the space of anti-symmetric functions in $C^\infty(\RRR^{Nd})$; likewise for other function spaces.

Let $L^2(F)=L^2(\nrd,F)$ denote the space of square-integrable (non-smooth) cross-sections of $F$. In more detail, we define $L^2(\Q,E)$ for any Hermitian bundle $E$ over any Riemannian manifold $\Q$ as follows (in particular, for the fermionic line bundle, $E=F$ and $\Q=\nrd$). We denote the Riemannian volume measure of $\Q$ simply by $dq$. For any $q\in \Q$, let $\inpr{\cdot}{\cdot}_q$ denote the inner product in the fiber space $E_q$. Consider a cross-section $\psi$ of $E$ that is not necessarily smooth but measurable, and call it square-integrable if
\be\label{sqint}
\int_{\Q} dq \, \inpr{\psi(q)}{\psi(q)}_q < \infty\,.
\ee
$L^2(\Q,E)$ is the space of all square-integrable cross-sections of $E$ modulo equality almost everywhere (relative to $dq$). It is a Hilbert space with the following inner product:
\be\label{scpL^2F}
\scp{\phi}{\psi} = \int_\Q dq \,\inpr{\phi(q)}{\psi(q)}_q\,.
\ee

\newcounter{propsmooth}
\setcounter{propsmooth}{\theprop}
\begin{prop}\label{prop:smooth}
For any $N\in\NNN$, $d\geq 2$, and any global trivialization $\hat{I}:\wh{F} \to \cid \times \CCC$, 
Equation \eqref{whpsihatI} defines a unitary isomorphism $U: L^2(F) \to \Anti L^2(\RRR^{Nd})$.
\end{prop}

The Hilbert space $\Anti L^2(\RRR^{Nd})$, of anti-symmetric square-integrable functions $\RRR^{Nd}\to \CCC$, is exactly the space usually utilized for representing fermionic quantum states. So we have a correspondence between that space and $L^2(F)$, and indeed the correspondence is canonical up to a global phase factor arising from the freedom in choosing the global trivialization $\hat{I}$. 

Via this correspondence $U$, we claim that the quantum theory for which wave functions are cross-sections of $F$ is equivalent to that for which wave functions are anti-symmetric functions on $\RRR^{Nd}$. It is not obvious how to make this claim mathematically precise; we do this in the following way: First, we show that $U$ maps the Schr\"odinger Hamiltonian on $L^2(F)$ to the usual Hamiltonian $H_{\Anti}$ on $\Anti L^2(\RRR^{Nd})$. More precisely, let us write $\tilde{H}_F$ for the operator
\be\label{rawH}
-\tfrac{\hbar^2}{2m} \Laplace + V\,,
\ee
defined on $C^\infty_0(F)$, the space of smooth cross-sections of $F$ with compact support; $\Laplace$ is the appropriate Laplace operator (based on the connection of $F$ and on the Laplace--Beltrami operator on the Riemannian manifold $\nrd$, see Section~\ref{sec:anti} for more detail); the potential $V:\RRR^{Nd}\to \RRR$ is assumed to be measurable, bounded,\footnote{Weaker assumptions on $V$ are sufficient; see Section~\ref{sec:anti} for more detail.} and symmetric under permutations,
\be\label{Vsymm}
V(\canq_1,\ldots,\canq_N) = V(\{\canq_1,\ldots,\canq_N\})\,.
\ee
Likewise, let $\tilde{H}_{\Anti}$ denote the expression \eqref{rawH}, understood as an operator defined on $\Anti C^\infty_0(\RRR^{Nd})$, with the usual Laplacian on $\RRR^{Nd}$, $\Laplace=\sum_{i=1}^{Nd} \partial^2/\partial x_i^2$, and the same function $V$. It is well known (for more detail, see the proof of Proposition~\ref{prop:HUHU} in Section~\ref{sec:anti}) that there is a unique self-adjoint extension $H_{\Anti}$ of $\tilde{H}_{\Anti}$ in $\Anti L^2(\RRR^{Nd})$.

\newcounter{propHUHU}
\setcounter{propHUHU}{\theprop}
\begin{prop}\label{prop:HUHU}
There is a self-adjoint extension $H_F$ of $\tilde{H}_F$ that satisfies
\be\label{Udomain}
U(\domain(H_F)) = \domain (H_{\Anti})\,,
\ee
where $\domain(H)$ means the domain of the operator $H$, and
\be\label{HUHU}
H_F = U^{-1}H_{\Anti} U\,.
\ee
As a consequence, $U$ intertwines the time evolution,
\be\label{TUTU}
e^{-iH_Ft/\hbar} = U^{-1} e^{-iH_{\Anti} t/\hbar} U\,.
\ee
\end{prop}

We note that $H_F$ is, in fact, the only self-adjoint extension of $\tilde{H}_F$ in $L^2(F)$ for $d\geq 4$ (and, we believe, also for $d=2,3$); for more detail, see the remarks in Section~\ref{sec:anti} after the proof of Proposition~\ref{prop:HUHU}.

\newcounter{propQUQU}
\setcounter{propQUQU}{\theprop}
\begin{prop}\label{prop:QUQU}
$U$ maps the position observable $\pvm_F$ to the position observable $\pvm_{\Anti}$,
\be\label{QUQU}
\pvm_F = U^{-1} \pvm_{\Anti} U\,.
\ee
\end{prop}

Here, we take the position observables (or rather, configuration observables) $\pvm_F$ and $\pvm_{\Anti}$ to be represented by projection-valued measures (PVMs); $\pvm_F$ is the obvious choice of PVM on $\nrd$ acting on $L^2(F)$, given by
\be\label{QFdef}
\bigl(\pvm_F(A) \psi\bigr)(q) = 1_A(q) \, \psi(q)
\ee
for any measurable set $A\subseteq \nrd$, with $1_A$ the indicator function; $\pvm_{\Anti}$ is the \emph{unordered} configuration observable that does not convey information about any particle's label (after all, labels cannot be measured), i.e., it is the PVM on $\nrd$ acting on $\Anti L^2(\RRR^{Nd})$ given by
\be\label{QAntidef}
\bigl(\pvm_{\Anti}(A)\wh{\psi}\bigr)(\covq) = 1_{\proj^{-1}(A)}(\covq) \, \wh{\psi}(\covq) 
\ee
for any measurable $A\subseteq \nrd$ and $\wh{\psi}\in \Anti L^2(\RRR^{Nd})$.

Together, Propositions \ref{prop:HUHU} and \ref{prop:QUQU} yield

\begin{prop}\label{prop:sim}
\be\label{equi}
\Bigl(L^2(F),H_F,\pvm_F \Bigr) \sim 
\Bigl( \Anti L^2(\RRR^{Nd}), H_{\Anti}, \pvm_{\Anti} \Bigr)\,.
\ee
\end{prop}

As a corollary of \eqref{thesis} and \eqref{equi}, we obtain the main claim of this paper: the quantum theory defined in terms of the fermionic line bundle $F=F(N,d)$, encoded in the quantum triple
$\bigl(L^2(F),H_F,\pvm_F \bigr)$,
is the quantum theory of $N$ fermions in $\RRR^d$.

\subsection{Examples of Equivalent Quantum Triples}

To further illustrate the concept of equivalence of quantum triples, we give some more examples.

\bigskip

\begin{ex}
A bosonic wave function (of $N$ spinless identical particles) can be represented either as a symmetric function on $\RRR^{Nd}$ (which is the standard choice) or as a function on $\nrd$. We base this claim on the equivalence
\be\label{bosonequi}
\Bigl(L^2(\nrd),H_B,\pvm_B\Bigr) \sim
\Bigl(\Sym L^2(\RRR^{Nd}), H_{\Sym},\pvm_{\Sym} \Bigr)\,.
\ee
Here, $H_B$ is a suitable self-adjoint extension of $-\frac{\hbar^2}{2m}\Delta +V$ with $\Delta$ the Laplace--Beltrami operator on $\nrd$ and $V\in L^\infty(\nrd)$ real-valued; $\pvm_B$ is defined in the same way as $\pvm_F$ in \eqref{QFdef}; ``$\Sym$'' is the symmetrization operator
\be
\Sym \wh{\psi}(\vx_1,\ldots,\vx_N) = \frac{1}{N!} \sum_{\sigma\in S_N} \wh{\psi}(\vx_{\sigma(1)},\ldots,\vx_{\sigma(N)}) \, ;
\ee
$H_{\Sym}$ is the unique self-adjoint extension of $-\frac{\hbar^2}{2m}\Delta+V$ with $\Delta=\sum_{i=1}^{Nd} \partial^2/\partial x_i^2$ the Laplacian on $\RRR^{Nd}$ and $V(\vx_1,\ldots,\vx_N)=V(\{\vx_1,\ldots,\vx_N\})$; and $\pvm_{\Sym}$ is defined in the same way as $\pvm_{\Anti}$ in \eqref{QAntidef}. The unitary isomorphism $U:L^2(\nrd)\to\Sym L^2(\RRR^{Nd})$ providing the equivalence is defined by the equation, for any $\psi\in L^2(\nrd)$,
\be
(U\psi)(\vx_1,\ldots,\vx_N) = \frac{1}{\sqrt{N!}} \psi(\{\vx_1,\ldots,\vx_N\})
\ee
for $(\vx_1,\ldots,\vx_N)\in \cid$. (We need not define $U\psi$ on $\RRR^{Nd}\setminus \cid$, the set of collision configurations, as that is a null set.)  Proofs very similar to those of Propositions~\ref{prop:HUHU} and \ref{prop:QUQU} (only simpler) show that $U$ maps $H_B$ to $H_{\Sym}$ and $\pvm_B$ to $\pvm_{\Sym}$.
\end{ex}

\begin{ex}
Suppose that $\gencon$ is any Riemannian manifold and that $E$ and $E'$ are two isomorphic Hermitian bundles over $\gencon$; then the quantum triples they give rise to are equivalent,
\be\label{EE'equi}
\Bigl( L^2(\gencon,E),H_E,\pvm_E \Bigr) \sim \Bigl( L^2(\gencon,E'),H_{E'}, \pvm_{E'} \Bigr)\,.
\ee
In detail, for any Hermitian bundle $E$ over $\gencon$ there is a natural PVM $\pvm_E$ on $\gencon$ acting on $L^2(\gencon,E)$, defined in the same way as $\pvm_F$ before, namely by
\be\label{QEdef}
\pvm_E(A) \psi (q) = 1_A(q) \, \psi(q)
\ee
for every measurable subset $A\subseteq \gencon$. Let $V:\gencon\to\RRR$ be a measurable function and $\Laplace$ the Laplace--Beltrami operator defined by the connection of $E$, and suppose that $-\tfrac{\hbar^2}{2m}\Laplace +V$ possesses a unique self-adjoint extension $H_E$.  Then, with $\Laplace'$ the Laplace--Beltrami operator of $E'$ and $I:E\to E'$ an isomorphism of Hermitian bundles, also $-\tfrac{\hbar^2}{2m}\Laplace'+V$ possesses a unique self-adjoint extension $H_{E'}$, and the equivalence \eqref{EE'equi} is realized by the unitary isomorphism $U:L^2(E) \to L^2(E')$ given by
\be
(U\psi)(q) = I_q \psi(q)\,.
\ee
\end{ex}

\begin{ex}
For a system of particles in a magnetic field, a gauge transformation of the electromagnetic vector potential leads to an equivalent quantum triple; this fact fits well with our thesis that equivalent quantum triples represent different mathematical descriptions of the same physical situation. For the sake of simplicity, consider a system consisting of a single spinless particle; the relevant equivalence reads
\be\label{gaugeequiv}
\Bigl(L^2(\RRR^3), H_{\vA}, \pvm \Bigr) \sim
\Bigl( L^2(\RRR^3), H_{\vA+\nabla f}, \pvm \Bigr)\,.
\ee
Here, the vector potential $\vA$ is a vector field on $\RRR^3$ (say, bounded and $C^1$ with bounded first derivatives); the function $f:\RRR^3\to\RRR$ (say, bounded and $C^2$ with bounded derivatives) defines the change of gauge that replaces $\vA$ by $\vA+\nabla f$; the Hamiltonian $H_{\vA}$ is the unique self-adjoint extension of $\frac{1}{2m}(-i\hbar\nabla - \vA)^2$; and $\pvm$ is the natural PVM as defined in \eqref{QEdef}. The unitary operator $U=U_f$ on $L^2(\RRR^3)$ providing the equivalence \eqref{gaugeequiv} is given by
\be
\bigl(U_f\psi\bigr)(\vx) = e^{if(\vx)/\hbar} \, \psi(\vx)\,.
\ee
Indeed, a simple calculation shows that $U_f$ maps $-i\hbar\nabla - \vA$ to $-i\hbar\nabla -( \vA+\nabla f)$ and thus $H_{\vA}$ to $H_{\vA+\nabla f}$; since $U_f$ is a multiplication operator, it commutes with $\pvm$ and thus maps $\pvm$ to itself; this establishes \eqref{gaugeequiv}.
\end{ex}

\subsection{Fermionic Fock Space}
\label{sec:Fock}

The \emph{bosonic Fock space} $\Fock^+$ and the \emph{fermionic Fock space} $\Fock^-$ for spinless particles in $\RRR^d$ are defined as
\begin{align}
\Fock^+ &= \bigoplus_{N=0}^\infty \Sym \, \bigl(L^2(\RRR^{d})^{\otimes N}\bigr)\\
\Fock^- &= \bigoplus_{N=0}^\infty \Anti \, \bigl(L^2(\RRR^{d})^{\otimes N}\bigr)\,.
\end{align}
The natural configuration space of a variable number of identical particles in $\RRR^d$ is
\be\label{Gammadef}
\Gamma(\RRR^d)= \Gamma = \{S\subset \RRR^d| \# S<\infty\}
=\bigcup_{N=0}^\infty \nrd\,.
\ee
Since each sector $\nrd$ is equipped with a measure $\mu_N$, the Riemannian volume measure, also $\Gamma$ is naturally equipped with a measure,
\be\label{mudef}
\mu(A) = \sum_{N=0}^\infty \mu_N(A\cap \nrd)
\ee
for all measurable sets $A\subseteq \Gamma$. 

The obvious Hilbert space associated with $\Gamma$ is $L^2(\Gamma,\mu)$, which coincides with the bosonic Fock space, 
\be
\Fock^+ \cong L^2(\Gamma,\mu)\,.
\ee
That is, there is a canonical unitary isomorphism $U:L^2(\Gamma,\mu) \to \Fock^+$, given by
\be
U\psi = \bigoplus_{N=0}^\infty \frac{1}{\sqrt{N!}}\psi\big|_{\nrd} \circ \proj\,.
\ee
For given $\psi:\Gamma\to\CCC$ one obtains the $N$-particle sector of $U\psi$ by
\be
(U\psi)(\canq_1,\ldots,\canq_N)=N!^{-1/2} \psi(\{\canq_1,\ldots,\canq_N\})
\ee
except on the null set of collision configurations. 

Since on each sector $\nrd$ of $\Gamma(\RRR^d)$ there is defined a fermionic line bundle $F(N,d)$, the (disjoint) union
\be
F(\Gamma(\RRR^d)) = F(\Gamma) = \bigcup_{N=0}^\infty F(N,d)
\ee
is a Hermitian line bundle over $\Gamma$, the \emph{fermionic line bundle over $\Gamma$}. The square-integrable cross-sections of this bundle relative to the measure \eqref{mudef} form a Hilbert space $L^2(\Gamma,F(\Gamma),\mu)$, which coincides with the fermionic Fock space, 
\be
\Fock^- \cong L^2(\Gamma, F(\Gamma),\mu)\,.
\ee
That is, there is an ``almost canonical'' unitary isomorphism $U: L^2(\Gamma,F(\Gamma),\mu) \to \Fock^-$, given by
\be
U\psi = \bigoplus_{N=0}^\infty U_N \psi\big|_{\nrd}\,,
\ee
where $U_N$ is the unitary isomorphism $L^2(F(N,d))\to \Anti L^2(\RRR^{Nd})$ provided by Proposition~\ref{prop:anti}. When saying that $U$ is ``almost canonical'' we mean that the only freedom in the choice of $U$ is the choice of one phase factor $e^{i\theta_N}$ for each $N=0,1,2,3,\ldots$.

An advantage of the possibility to write $\Fock^-$ as an $L^2$ space is that the probability distribution on $\Gamma$ that a Fock state $\psi\in\Fock^-$ with $\|\psi\|=1$ defines has density immediately given by $|\psi|^2$.

As a last remark, it seems even more persuasive in connection with Fock spaces than with a fixed number $N$ of particles that unordered configurations are more natural than ordered ones. For example, if, in the ordered 3-particle configuration $(\vx_1,\vx_2,\vx_3)$, particle 2 gets annihilated, then particle 3 has to be renumbered as particle 2, and it seems clear that this change in numbering does not correspond to any change in the physical state in nature. Likewise, when a new particle gets created, which number should it get?

\bigskip

This completes our overview of the fermionic line bundle. We now turn to a discussion of the details.

\section{Constructions and Properties of the Fermionic Line Bundle}
\label{sec:fermionicbundle}

\subsection{Constructions}

The proof of Proposition~\ref{prop:representation} provides a construction of a Hermitian bundle with a desired holonomy representation, and thus in particular a construction of the fermionic line bundle. We now describe three further constructions of the fermionic line bundle.

\begin{itemize}
\item Our first construction of $F(N,d)$ utilizes the \emph{set-indexed tensor product} that we discuss in detail elsewhere \cite{GTTZ14b}. In contrast to the usual tensor product $W_1\otimes W_2\otimes \cdots \otimes W_N$, where the factors are ordered, the set-indexed tensor product $\bigotimes_{a\in T} W_a$ uses an arbitrary finite (unordered) index set $T$. Correspondingly, one can form the set-indexed tensor power $W^{\otimes T}$ by introducing a copy $W_a$ of $W$ for every $a\in T$ and then forming their set-indexed tensor product. Now let $W$ be an arbitrary finite-dimensional Hermitian vector space (i.e., Hilbert space) and use unordered configurations $q$ as the index set $T$. Let $E$ be the bundle $\cup_{q\in\Q} W^{\otimes q}$; it naturally inherits the structure of a Hermitian bundle \cite{GTTZ14b}. Let $E'$ be the
subbundle of totally anti-symmetric elements.  The difference between
$E'_q$ and the $N$-th exterior power of $W$, $\Lambda^N W \subset
W^{\otimes N}$, is that the relevant permutations are those of $q$
instead of those of $\{1, \ldots, N\}$. $E'_q$ could thus be denoted
$\Lambda^q W$; it is relevant here that $W$ is always the same space,
and not a bundle $W_{\canq}$, since in contrast to tensor products one
cannot form ``exterior products'' of several spaces, only exterior
powers of the same space.  The dimension of $E'_q$ is the binomial
coefficient $\binom{\dim W}{N}$. 
Having thus defined the fiber space $E'_q$ at every $q$, we
obtain a Hermitian inner product from that of $E_q$, and a connection
from that of $E$ using the fact that $E'$ is a \emph{parallel}
subbundle of $E$, i.e., parallel transport from $E_q$ to $E_r$ maps
$E'_q$ to $E'_r$. Hence, $E'$ is a \herm\ bundle. Finally, assume that
$W$ has dimension $N$ equal to the number of particles, and set $F =
E'$.

To see that $F$ is the fermionic line bundle, note first that $F$ is a
\herm\ bundle and has rank 1.  Furthermore, parallel transport
around a loop $\closedcurve$ permutes the factors, so that (in an
obvious notation borrowed from exterior products)
\begin{equation}
  \lombdo_{\gena \in \genset} \lilgenvec_{\gena}
  \overset{\holonomy_\closedcurve}{\longmapsto} \lombdo_{\gena \in
  \genset}{\lilgenvec_{\deckt\gena}} = (-1)^{\sigma} \lombdo_{\gena
  \in \genset} \lilgenvec_{\gena}
\end{equation}
where each $w_{\canq}$ is an element of $W$, and
$\holonomy_\closedcurve$ is the holonomy $E_q \to E_q$.

\item The second construction of $F = F(N,\dd)$ works only for odd
$\dd$: the fiber $F_q$ is the space of complex pseudo-scalars of the
tangent space $T_q \Q$ of $\Q=\nrd$,
\begin{equation}\label{Fconstr1}
  F_q = \CCC \Lambda^{\! N\dd} \,T_q \Q\,,
\end{equation}
where $\Lambda^k \Euclideanspace$ denotes the $k$-th exterior power of
the Euclidean vector space $\Euclideanspace$.  Hence, a cross-section of $F$
is a complex-valued differential form\footnote{A \emph{differential
form} is a tensor field that is completely anti-symmetric against
permutation of indices.  We utilize here that the cotangent space
$T^*_q \Q$ is canonically identified with the tangent space $T_q \Q$
by means of the metric.} of maximal degree $N\dd$ over $\Q = \nrd$.
The complexification, written here by a prefix $\CCC$, is to be understood as turning a Euclidean
vector space into a Hermitian vector space. The connection of $F$ is
inherited from the connection of the tangent bundle $T\Q$.

To see that \eqref{Fconstr1} agrees with the definition of the fermionic
line bundle, we have to compute the holonomy for an arbitrary loop
$\closedcurve$ starting and ending at $q$. Choose an ordering of the
points in $q$, say $q = \{ \canq_1, \ldots, \canq_N\}$, and an
orthonormal basis $\ve_1, \ldots, \ve_d$ of $\RRR^\dd$. We thus have
an (ordered) orthonormal basis of $T_q\Q = \oplus_{\canq \in q} T_{\canq}
\RRR^\dd$ (namely $\ve_{1,1}, \ldots, \ve_{1,\dd}, \ldots, \ve_{N,1},
\ldots, \ve_{N,\dd}$), which defines a ``basis'', i.e., an element of
norm one, of $F_q$, namely $\omega := \ve_{1,1} \wedge \cdots \wedge
\ve_{1,\dd} \wedge \cdots \wedge \ve_{N,1} \wedge \cdots \wedge
\ve_{N,\dd}$ with $\wedge$ the exterior product.  Parallel transport
along $\closedcurve$ leads to $\holonomy_\closedcurve \omega =
\ve_{\sigma(1),1} \wedge \cdots \wedge \ve_{\sigma(1),\dd} \wedge
\cdots \wedge \ve_{\sigma (N),1} \wedge \cdots \wedge \ve_{\sigma
(N),\dd}$ where $\sigma \in S_N$ represents the permutation of $q$
carried out by $\closedcurve$ in terms of the ordering of $q$ chosen
above. Since $\holonomy_\closedcurve \omega$ differs from $\omega$
just by the ordering of the factors, it differs only by a sign. This
sign is, in fact, $+1$ if $\dd$ is even and $(-1)^\sigma$ if $\dd$ is
odd, which completes the proof.

\item Our third construction begins with defining a
\herm\ vector bundle $E$ of rank $N!$ by
\begin{equation}\label{idenEdef}
   E_q := \bigoplus_{\covq \in \proj^{-1}(q)} \CCC\,,
\end{equation}
where $\pi:\cid\to\nrd$ maps any ordered configuration to the corresponding unordered one,
\begin{equation}
\pi(\vx_1,\ldots,\vx_N) = \{\vx_1,\ldots,\vx_N\}\,,
\end{equation}
so $\covq$ runs through all possible orderings of the unordered configuration $q$;
the set-indexed direct sum is to be understood as an orthogonal sum; and the
connection is the obvious one: parallel transport along a curve
$\opencurve$ from $q$ to $r$ maps $w \in E_q$ to the $w' \in E_r$
having components $w'_{\hat{r}} = w_{\hat{q}}$ where $\hat{r}$ is the
endpoint of the lift of $\opencurve$ to $\cid$ starting from
$\hat{q}$.  Every function $\psi:\cid \to \CCC$ naturally gives
rise to a cross-section $\phi$ of $E$, defined by
\begin{equation}
 \phi(q) = \oplos\limits_{\covq
  \in \proj^{-1}(q)} \psi(\covq).  
\end{equation}
The anti-symmetric functions on
$\cid$ then all correspond to cross-sections lying within a rank-1
subbundle $F$ of $E$, defined by
\begin{equation}\label{FsubidenE}
  F_q = \{ w \in E_q | w_{\deckt \covq} = (-1)^{\deckt} w_{\covq}
  \: \forall \deckt \} \,.
\end{equation}
Having thus defined the fiber space $F_q$ at every $q$, we obtain a
Hermitian inner product from that of $E_q$, and a connection from that
of $E$ using the fact that $F$ is a \emph{parallel} subbundle of $E$.

To see that $F$ is the fermionic line bundle, we have to compute the
holonomy for an arbitrary loop $\closedcurve$ starting and ending at
$q$. {}From the definition of the connection we have that, if the lift
of $\closedcurve$ starting at $\covq_0$ leads to $\covq = \deckt
\covq_0$, $(\holonomy_\closedcurve w)_{\covq} = w_{\covq_0} =
w_{\deckt^{-1} \covq} = (-1)^{\deckt} w_{\covq}$, which is what we
needed to show.
\end{itemize}

\subsection{Remarks}
\label{sec:rem}

\begin{enumerate}

\item Nontriviality. The fermionic line bundle for $d \geq 3$ possesses not only a
  nontrivial connection, it is also nontrivial as a vector bundle
  (disregarding connection and Hermitian inner products). To see this,
  consider the trivial \herm\ bundle $\nrd \times \CCC$.  Note that
  any two connections on the same line bundle differ by a complex
  1-form $A$ in the sense $\nabla_\mu' = \nabla_\mu + A_\mu$. It
  follows that holonomies differ by the integral of $A$ over the loop
  in the sense $\holonomy_\closedcurve' = \exp(\int_\closedcurve A)
  \holonomy_\closedcurve$. If both connections are flat, $A$ must be
  closed, $dA=0$. Now for $d\geq 3$, all closed 1-forms $A$ on $\nrd$
  are exact, i.e., $A=df$ for some function $f:\nrd\to\CCC$; in
  other words, $\nrd$ has trivial first cohomology group. 
  
  To see this,
  note that every closed 1-form on a manifold $\Q$, when lifted to a 
  1-form $\hat{A}$ on the (simply connected) universal covering space 
  $\wh{\Q}$, becomes an exact 1-form, $\hat{A}=d\hat{f}$. Since every
  deck transformation $\sigma$ of $\wh{\Q}$ will carry $\hat{A}$ to itself, it will
  carry $\hat{f}$ to $\hat{f}+\gamma_\sigma$, where $\gamma_\sigma\in \CCC$
  is some constant. Therefore, $\gamma_{\sigma\circ \sigma'} =
  \gamma_\sigma + \gamma_{\sigma'}$ and $\gamma_{\id}=0$;
  in other words, $\gamma$ is a homomorphism from the group of 
  deck transformations of $\wh{\Q}$ over $\Q$ to the additive group of
  the complex numbers. The group of deck transformations is isomorphic to the 
  fundamental group of $\nrd$, which for $d\geq 3$ is isomorphic to
  the permutation group $S_N$ of $N$ objects, and thus is finite.
  However, there is no nontrivial homomorphism from a finite group to 
  the additive group of the complex numbers
because every element $\sigma$ of the group has finite order,
$\sigma^n = \id$, so that $n \gamma_\sigma =\gamma_{\sigma^n}=0$ and thus
$\gamma_{\sigma} =0$.
  
  So we have shown that $A$ is exact, $A=df$.
  But then $\int_\closedcurve A =
  0$, therefore $\holonomy_\closedcurve' = \holonomy_\closedcurve$ so
  that all holonomies $\holonomy_\closedcurve'$ must be trivial.
  Therefore, no choice of connection and Hermitian inner products can
  make the trivial line bundle over $\nrd$ a fermionic line bundle.

 The same proof shows, in fact, that any flat Hermitian line bundle with
 non-trivial holonomy over a base manifold with finite fundamental group
 cannot be trivialized as a vector bundle. For example, the restriction of 
 the fermionic line bundle $F(N,d)$ to an open subset $\Q$ of $\nrd$, or to
 a submanifold $\Q$ of $\nrd$, such that the fundamental group of $\Q$ is finite
 and the holonomy non-trivial,
 is non-trivial as a vector bundle.

\item Nodes. The fact that $F$ is non-trivial as a vector bundle implies that
  every (smooth) cross-section has nodes, i.e., it vanishes somewhere. That is
  because any nowhere-vanishing (smooth) cross-section of a line bundle defines a
  trivialization of the bundle. As a consequence, every smooth
  anti-symmetric function on $\rdn$ has nodes outside the ``diagonal'' (i.e.,
  the set $\RRR^{Nd}\setminus \cid$ of collision configurations).
  Another consequence is that $\nrd$ is not
  orientable for odd $d$. That is because every orientable manifold
  permits a volume form, i.e., a real-valued, nowhere-vanishing smooth
  differential form of maximal degree, which would provide a
  particular case of a complex-valued nowhere-vanishing smooth form of degree
  $Nd$ and thus, according to the second construction of $F$, a
  nowhere-vanishing smooth cross-section of $F$. 
  Also, by considering suitable neighborhoods near the boundary, we can make a similar
argument to deduce that a smooth cross-section of $F$ has zeros arbitrarily close to any point on the
boundary.

\item\label{sec:idencurved}
Fermions in Curved Space.
The fermionic line bundle can be defined in the same way if physical space is curved. (For example, this case arises in curved space-time, wherein physical space corresponds to a spacelike hypersurface.) In this case, we take physical space to be mathematically represented by a Riemannian manifold $M$ of dimension $d\geq 2$. Then the configuration space of $N$ identical particles is the set of $N$-element subsets of $M$, denoted by \nm{M}. It is canonically identified with $M^N_{\neq}/S_N$, where
\begin{equation}
  M^N_{\neq} = \bigl\{(\vx_1, \ldots, \vx_N) \in M^N \big| 
  \vx_j \neq \vx_k \:\:\forall j \neq  k \bigr\}.
\end{equation}
By means of this identification, $\nm{M}$ obtains the structure of a Riemannian
manifold; it has dimension $Nd$. 

The fermionic line bundle on $\nm{M}$, $F = F(N,M)$ is defined to be \emph{the 
Hermitian line bundle over $\nm{M}$ such that the holonomy associated
with a loop $\closedcurve$ in $\nm{M}$ starting and ending at $q$ is
$\holonomy_\closedcurve = (-1)^{\sigma}$ where $\sigma$ is the
permutation of $q$ defined by $\closedcurve$.} By Proposition~\ref{prop:representation} (applied to every connected component of $\nm{M}$ separately), $F(N,M)$ exists and is unique up to isomorphism; also the three further constructions work in the same way (one of them for odd $d$ only). Despite the curvature of $M$, $F(N,M)$ is flat; a related fact is that for any (curved) Riemannian
manifold $\Q$ the bundle of pseudo-scalars, $\Lambda^{\dim \Q} T \Q$, is flat.

\item\label{rem:anyons} Anyons. The manifold ${}^N\RRR^2$ differs notably from $\nrd$ for $d\geq 3$ in a certain topological respect: its fundamental group $\pi_1({}^N \RRR^2)$ is isomorphic to the braid group of $N$ strings, whereas the fundamental group of $\nrd$ for $d\geq 3$ is the permutation group $S_N$ of $N$ objects. It has been argued that the possible quantum theories on a manifold $\Q$ are in one-to-one correspondence with the characters (i.e., 1-dimensional unitary representations) $\gamma$ of the fundamental group $\pi_1(\Q)$ \cite{LM77,topid0,topid1A}. While $S_N$ has two characters ($\gamma_\sigma =1$ and $\gamma_\sigma = (-1)^\sigma$, corresponding to bosons and fermions), the braid group has a 1-parameter family of characters, $\gamma^{(\beta)}$ with parameter $\beta\in \RRR$, namely $\gamma_\sigma^{(\beta)}=e^{i\beta}$ whenever $\sigma$ is the exchange of two particles. Bosons correspond to $\beta = 0$ and fermions to $\beta = \pi$.  The
other possibilities are called \emph{fractional statistics}, and the
corresponding particles \emph{anyons}. They were first suggested in
\cite{LM77}, and their investigation began in earnest with
\cite{GMS81, Wi82}; see \cite{Mor92} for some more details and
references. For each value of $\beta$, the corresponding anyonic wave functions can be represented by cross-sections of an \emph{anyonic line bundle} $A^{(\beta)}$ over ${}^N\RRR^2$, defined by the property that \emph{the holonomy associated with a loop $\closedcurve$ in ${}^N\RRR^2$ starting and ending at $q$ is $\holonomy_\closedcurve = \gamma^{(\beta)}_{[\closedcurve]}$, where $[\closedcurve]$ is homotopy class of $\closedcurve$.}

\item Dimension $d=1$. The manifold $\RRR^{1,N}_{\neq}$ is not connected but has $N!$ connected components, each of which is isometric to ${}^N\RRR^1$. The latter space is simply connected, so that every flat Hermitian bundle over it is trivial; as a consequence, the definition of the fermionic line bundle, while still meaningful for $d=1$, yields only a trivial Hermitian bundle as $F(N,1)$. Thus, for $d=1$ the difference between fermions and bosons is not captured in a bundle. In fact, for $d=1$ the operator $\tilde{H}_F$ considered in Proposition~\ref{prop:HUHU} and defined in \eqref{rawH} possesses more than one self-adjoint extension, corresponding to different boundary conditions on the boundary of ${}^N\RRR^1$ (or, equivalently, on the collision configurations in $(\RRR^1)^N$); we expect that with Dirichlet boundary conditions the quantum triple is equivalent to fermions, with Neumann boundary conditions to bosons. For $d\geq 4$ (and presumably also for $d=2,3$), in contrast, the fermionic line bundle permits only Dirichlet boundary conditions (as follows from the uniqueness of the self-adjoint extension of $\tilde{H}_F$).

\end{enumerate}

\section{Proofs of Propositions \ref{prop:hatFtrivial}--\ref{prop:QUQU}} 
\label{sec:anti}

\subsection{Equivalence to Anti-Symmetric Wave Functions}

\setcounter{merken}{\theprop}
\setcounter{prop}{\theprophatFtrivial}
\begin{prop}
For any $N\in\NNN$ and $d \geq 2$, the lift $\wh{F} = \proj_* F$ of the fermionic line bundle to $\cid$ is a trivial Hermitian bundle.
\end{prop}
\setcounter{prop}{\themerken}

\begin{myproof}
For $d\geq 3$, $\cid$ is the universal covering space of $\nrd$, and the proof of Proposition~\ref{prop:representation} in Appendix~\ref{app:representation} shows that $\wh{F}$ is a trivial Hermitian bundle. For $d=2$, $\cid$ is \emph{a} covering space of $\nrd$, but not the universal covering space; that is, $\cid$ is not simply connected. For this reason, we include the following proof of Proposition~\ref{prop:hatFtrivial}, which works in any dimension $d\geq 2$.

The lift $\wh{E}=\proj_*(E)$ of a Hermitian bundle $E$ over $\nrd$ to $\cid$ can be constructed by setting, for every $\covq\in\cid$, $\wh{E}_{\covq} = E_{\proj(\covq)}$. 
A Hermitian bundle is trivial if and only if all holonomies are. (Indeed, a trivial Hermitian bundle necessarily has trivial holonomies; a Hermitian bundle $E$ with trivial holonomies can be trivialized by choosing a point $q$ in the base manifold and an orthonormal basis in $E_q$ and parallel transport of this basis to every other point $q'$ along any curve; the resulting orthonormal basis of $E_{q'}$ does not depend on the choice of the curve because otherwise the loop obtained by concatenation of one curve and the inverse of the other would have nontrivial holonomy.) 

Thus, it suffices to show that $\wh{F}$ has trivial holonomies. So consider a loop $\closedcurve$ in $\cid$ based at $\covq\in\cid$ and observe that its projection $\proj\closedcurve$ to $\nrd$ is a loop based at $q=\proj(\covq)$ for which the corresponding permutation $\sigma_{\proj\closedcurve}$ of $q$ is the identity. Thus, by the definition of the fermionic line bundle $F$, $\proj\closedcurve$ has trivial holonomy. By construction of $\wh{F}$, the holonomy of $\closedcurve$ in $\wh{F}$ is the same as the holonomy of $\proj\closedcurve$ in $F$.
\end{myproof}

We also claimed that any two choices of isomorphisms of Hermitian bundles $\hat{I}:\wh{F}\to\cid\times\CCC$ differ at most by a global phase factor. This follows from the fact that any isomorphism-of-Hermitian-bundles $\hat{J}$ of the trivial Hermitian bundle $\cid\times\CCC$ to itself is just multiplication by a complex constant of modulus 1. (Indeed, $\hat{J}$ can be regarded as multiplication by a function $f:\cid\to\CCC$ whose values must have modulus 1 everywhere for the inner product to be preserved, and $\hat{J}$ maps the connection to itself if and only if $\nabla f=0$, which implies $f$ is constant since $\cid$ is connected.)

\setcounter{merken}{\theprop}
\setcounter{prop}{\thepropanti}
\begin{prop}
For any $N\in\NNN$, $d\geq 2$, any global trivialization $\hat{I}:\wh{F} \to \cid \times \CCC$, and any cross-section $\psi$ of $F$, $U\psi$ as given by 
\be\label{Uexplicit}
\bigl(U\psi\bigr)(\canq_1,\ldots,\canq_N) = 
\frac{1}{\sqrt{N!}}\hat{I}_{(\canq_1\ldots\canq_N)} \psi(\{\canq_1,\ldots,\canq_N\})
\ee
is an anti-symmetric function on $\cid$.
\end{prop}

\begin{myproof}
To see that $U\psi$ is an anti-symmetric function, let $\covq=(\canq_1,\ldots,\canq_N)\in\cid$, $q=\proj(\covq)\in\nrd$, and let $\sigma\in S_N$ be any transposition, exchanging two elements of $\{1,\ldots,N\}$, say $i$ and $j$. Let $\closedcurve$ be a loop in $\nrd$ based at $q$ such that the permutation of $q$ defined by $\closedcurve$ is the transposition of $\canq_i$ and $\canq_j$. By the definition of $F$, the holonomy $\holonomy_\closedcurve$ associated with $\closedcurve$ is $-1$. Since $\proj$ is a covering map, there is a unique continuous lift $\hat\closedcurve$ of $\closedcurve$ in $\cid$ starting at $\covq$; the curve $\hat\closedcurve$ is not closed but ends at
\be\label{sigmacovq}
\sigma \covq := (\canq_{\sigma (1)},\ldots,\canq_{\sigma (N)})\,.
\ee
Note that $\hat{I}$, the global Hermitian trivialization of $\wh{F}$, can be regarded as a cross-section of the dual bundle $\wh{F}^*$ that is parallel, $\nabla \hat{I}=0$, with $\nabla$ the covariant derivative of the connection of $\wh{F}^*$ dual to that of $\wh{F}$. Thus, $\hat{I}_{\sigma\covq}$ can be obtained from $\hat{I}_{\covq}$ by parallel transport in $\wh{F}^*$ along $\hat\closedcurve$; since parallel transport and lift commute, $\hat{I}_{\sigma \covq}$ can also be obtained from $\hat{I}_{\covq}$ by parallel transport in $F^*$ along $\closedcurve$, using that $\wh{F}_{\covq} = F_q$. But parallel transport along $\closedcurve$ leads to
\be\label{Itransposition}
\hat{I}_{\sigma \covq} = \hat{I}_{\covq} \circ \holonomy_{\closedcurve} = -\hat{I}_{\covq}\,,
\ee
so that
\be
\bigl(U\psi\bigr)(\sigma\covq) = 
\frac{1}{\sqrt{N!}}\hat{I}_{\sigma\covq} \psi(q)=
-\frac{1}{\sqrt{N!}}\hat{I}_{\covq} \psi(q)=
-\bigl(U\psi\bigr)(\covq)\,,
\ee
which means that $U\psi$ is anti-symmetric.
\end{myproof}

\setcounter{prop}{\thepropsmooth}
\begin{prop}
Equation \eqref{Uexplicit} defines a unitary isomorphism $U: L^2(F) \to \Anti L^2(\RRR^{Nd})$.
\end{prop}
\setcounter{prop}{\themerken}

\begin{myproof}
$\psi \mapsto U\psi$ preserves inner products because
\begin{align}
\int_{\cid} d\covq\, U\phi(\covq)^* \, U\psi(\covq) &=
\frac{1}{N!}\int_{\cid} d\covq\, \bigl(\hat{I}_{\covq} \phi(\proj\covq)\bigr)^* 
\hat{I}_{\covq} \psi(\proj\covq)\nonumber\\
&=\frac{1}{N!}\int_{\cid} d\covq\, \inpr{\phi(\proj\covq)}{\psi(\proj\covq)}_{\proj\covq}\nonumber\\
\label{tildeUscp}
&=\int_{\nrd} dq\, \inpr{\phi(q)}{\psi(q)}_q\,,
\end{align}
using that $\proj$ is an $N!$-to-one covering map that maps the Lebesgue measure $d\covq$ on $\cid$ to the Riemannian volume measure $dq$ on $\nrd$. As a consequence, if $\psi$ is square-integrable then $U\psi$ is square-integrable as well and, since the difference between $\RRR^{Nd}$ and $\cid$ is a null set, defines a square-integrable function on $\RRR^{Nd}$, which is anti-symmetric by Proposition~\ref{prop:anti}. 

We thus obtain a linear operator $L^2(F)\to \Anti L^2(\RRR^{Nd})$ as follows: for given $\psi\in L^2(F)$ let $\psi_0$ be any representative of the equivalence class that $\psi$ is, insert $\psi_0$ for $\psi$ on the right hand side of \eqref{Uexplicit} and obtain thus a function on $\cid$; define the function to be zero on the collision configurations $\RRR^{Nd}\setminus \cid$; take its equivalence class modulo equality almost everywhere and obtain thus an element of $L^2(\RRR^{Nd})$, which does not depend on the choice of $\psi_0$. Indeed, changing $\psi_0$ on a null set $A$ will change \eqref{Uexplicit} on the null set $\proj^{-1}(A)$, so \eqref{Uexplicit} does define a mapping $L^2(F)\to L^2(\RRR^{Nd})$.
Thus, $U$ is a linear operator $L^2(F)\to \Anti L^2(\RRR^{Nd})$ that preserves inner products. 

It remains to show that $U$ is surjective. Let $\wh{\psi}$ be an anti-symmetric, measurable function on $\cid$, let $q\in\nrd$ and $\covq,\covq' \in \proj^{-1}(q)$; then $\covq=(\canq_1,\ldots,\canq_N)$ and $\covq'=(\canq_1',\ldots,\canq_N')$ are just different orderings of the $N$ points in $\RRR^d$ collected in $q$. Thus, they differ by a permutation $\sigma$, $\canq'_i=\canq_{\sigma(i)}$ for all $i=1,\ldots,N$; in the notation \eqref{sigmacovq}, $\covq'=\sigma \covq$. Using \eqref{Itransposition} and the anti-symmetry of $\wh\psi$, we have that
\be\label{surjective}
\sqrt{N!} \, \hat{I}^{-1}_{\sigma\covq} \wh\psi(\sigma\covq) = 
\sqrt{N!} \, \hat{I}^{-1}_{\covq} \wh\psi(\covq) =: 
\psi(q)
\ee
consistently defines a measurable cross-section $\psi$ of $F$ such that $U\psi = \wh\psi$. Since $U$ preserves $L^2$ norms, $\psi$ is square-integrable if $\wh\psi$ is. Thus, $U$ is surjective and therefore unitary.
\end{myproof}

We also note for later use (in the proof of Proposition~\ref{prop:HUHU}) that \eqref{Uexplicit} also defines a bijective operator
$\tilde U:C^\infty(F)\to\Anti C^\infty(\cid)$. Indeed, note first that, since $\hat{I}$ is a diffeomorphism between the bundles $\wh{F}$ and $\cid \times \CCC$, $\tilde U$ maps any smooth cross-section $\psi$ of $F$ to a smooth and, by Proposition~\ref{prop:anti}, anti-symmetric function on $\cid$. To see that $\tilde U$ is injective, suppose that $\psi'\neq \psi$, so that $\psi'(q)\neq \psi(q)$ for some $q\in\nrd$; since for any $\covq \in\proj^{-1}(q)$, $\hat{I}_{\covq}$ is injective, $\tilde{U} \psi'(\covq) \neq \tilde{U}\psi(\covq)$. Now let $\wh{\psi}$ be an anti-symmetric, smooth function on $\cid$, let $q\in\nrd$ and $\covq,\covq' \in \proj^{-1}(q)$; then $\covq'=\sigma \covq$, and by the same reasoning as in \eqref{surjective} we obtain a smooth cross-section $\psi$ of $F$ (smooth because $\wh\psi$ and $\hat{I}$ are) such that $\tilde{U}\psi=\wh\psi$.

\bigskip

We are using the following definition of the gradient $\nabla$ and Laplacian $\Laplace$ on a Hermitian bundle $E$ over a Riemannian manifold $(\Q,g)$:
\begin{itemize}
\item By the gradient $\nabla f$ of a smooth
function $f: \Q \to \RRR$ we mean the tangent vector field on $\Q$
metrically equivalent (by ``raising the index'') to the 1-form $df$,
the differential of $f$. If $W$ is a complex vector space then, for a function $\psi: \Q \to W$, the differential $d\psi$ is a $W$-valued 1-form, and thus $\nabla \psi
(q) \in \CCC T_q \Q \otimes W$, where $\CCC T_q \gencon$ denotes
the complexified tangent space at $q$, and the tensor product
$\otimes$ is, as always in this paper, over the complex numbers.
When $\psi$ is a cross-section of a Hermitian bundle $E$, the covariant derivative $D\psi$ is an ``$E$-valued
1-form,'' i.e., a cross-section of $\CCC T\gencon^* \otimes E$ (with $T\Q^*$ the
cotangent bundle), while we write $\nabla \psi$ for the cross-section of $\CCC
T\gencon \otimes E$ metrically equivalent to $D\psi$.  

\item
The Laplace--Beltrami operator (or briefly, Laplacian) $\Laplace f$ of a function $f$ is defined to be the
divergence of $\nabla f$, where the divergence of a vector field $X$
is defined by
\begin{equation}
  \mathrm{div}\, X = D_a X^a
\end{equation}
(using the sum convention) with $D$ the (standard) covariant derivative operator, corresponding
to the Levi-Civita connection on the tangent bundle of $\Q$ arising
from the metric $g$. Equivalently, the divergence can be expressed in local coordinates as
\be
\mathrm{div}\, X = \frac{1}{\sqrt{\det g}} \partial_a \Bigl( \sqrt{\det g}\, X^a\Bigr)
\ee
with $\partial_a$ the partial derivative operator.
Since $D g =0$, we can write
\begin{equation}
  \Laplace f = g^{ab} D_a D_b f \,,
\end{equation}
where the second $D$, the one which is applied first, actually does
not make use of the Levi-Civita connection. In other words, the
Laplacian is the metric trace of the second (covariant) derivative.
Another equivalent definition is $\Laplace f = {*}d{*}d f$ 
where $d$ is the exterior derivative of differential forms and $*$ is the
Hodge star operator (see, e.g., \cite{EL83}).\footnote{The Hodge
  operator $*$ depends on the orientation of $\Q$ in such a way that a
  change of orientation changes the sign of the result. Thus, $*$ does
  not exist if $\Q$ is not orientable. However, it exists locally for
  any chosen local orientation, and since the Laplacian contains two
  Hodge operators, it is not affected by the sign ambiguity.} For
``$E$-valued functions'' $\psi$ (i.e., cross-sections of $E$) the Laplacian $\Laplace\psi$ is defined
correspondingly as the divergence of the ``$E$-valued vector
field'' $\nabla \psi$, or equivalently by
\begin{equation}
  \Laplace \psi = g^{ab} D_a D_b \psi 
\end{equation}
or by $\Laplace \psi = {*}d{*}d \psi$, using the obvious
extension of the exterior derivative to $E$-valued differential
forms.

To define the
covariant derivative of $D\psi$, one uses the connection on $\CCC
T\gencon^* \otimes E$ that arises in an obvious way from the Levi-Civita
connection on $\CCC T \gencon^*$ and the given connection on $E$, with the
defining property $D_{\CCC T \gencon^* \otimes E} (\omega \otimes \psi) =
(D_{\CCC T \gencon^*} \omega) \otimes \psi + \omega \otimes (D_E \psi)$ for
every smooth 1-form $\omega$ and every smooth cross-section $\psi$ of $E$. We take as the
Laplacian $\Laplace \psi$ of $\psi$  the (Riemannian)
metric trace of the second covariant derivative of $\psi$,
\begin{equation}\label{Laplacedef}
  \Laplace \psi = g^{ab} D_a D_b \psi\,, 
\end{equation}
where the second $D$, the one which is applied first, is the covariant
derivative on $E$, and the first $D$ is the covariant derivative on $\CCC
T\Q^* \otimes E$.\footnote{While this is the natural definition
  of the Laplacian of a cross-section of a \herm\ bundle, we note that for
  differential $p$-forms with $p \geq 1$ there are two inequivalent
  natural definitions of the Laplacian: one is $\Laplace = -(d^*d + dd^*)$
  (sometimes called the de Rham Laplacian, with $d^* = (-1)^{(\dim \Q)
    (p+1)+1} {*}d{*}$ on $p$-forms \cite[p.~9]{EL83}), the other is
  \eqref{Laplacedef} for $E = \CCC \Lambda^p T\Q^*$ (sometimes called the
  Bochner Laplacian). They differ by a curvature term given by the
  Weitzenb\"ock formula \cite[p.~11]{EL83}.}
Again, an equivalent definition is $\Laplace \psi = {*}d{*}d \psi$, using the
obvious extension (based on the connection of $E$) of the exterior
derivative to $E$-valued differential forms, i.e., cross-sections of $\CCC
\Lambda^p T \Q^* \otimes E$.
\end{itemize}

\setcounter{merken}{\theprop}
\setcounter{prop}{\thepropHUHU}
\begin{prop}
There is a self-adjoint extension $H_F$ of $\tilde{H}_F$ that satisfies
$U(\domain(H_F)) = \domain (H_{\Anti})$,
where $\domain(H)$ means the domain of the operator $H$, and
$H_F = U^{-1}H_{\Anti} U$.
As a consequence, $U$ intertwines the time evolution,
$e^{-iH_Ft/\hbar} = U^{-1} e^{-iH_{\Anti} t/\hbar} U$.
\end{prop}
\setcounter{prop}{\themerken}

\begin{myproof}
We first provide some detail about why $\tilde{H}_{\Anti}$ is essentially self-adjoint, i.e., possesses a unique self-adjoint extension $H_{\Anti}$ in $\Anti L^2(\RRR^{Nd})$. Let us write $\tilde{H}$ for the operator $-\tfrac{\hbar^2}{2m} \Laplace + V$ on $C_0^\infty(\RRR^{Nd})$. We first note the well-known fact \cite{RS75} that for bounded $V$, $\tilde{H}$ is essentially self-adjoint in $L^2(\RRR^{Nd})$. (In fact, this is the case for a much wider class of potentials $V$, as can be shown using the Kato--Rellich theorem \cite[Thm.~X.12]{RS75}. For example, for $d=3$ it is sufficient that
\be
V(\canq_1,\ldots,\canq_N) = V_{\mathrm{bdd}}(\canq_1,\ldots,\canq_N) +
\sum_{1\leq i<j\leq N} V_{\mathrm{pair}}(\canq_i-\canq_j)\,,
\ee
where $V_{\mathrm{bdd}}$ is bounded and the pair potential $V_{\mathrm{pair}}:\RRR^3\to\RRR$ is square-integrable \cite[Thm.~X.16]{RS75}; thus, the Coulomb potential is included.) 

Using that the potential $V$ is permutation-symmetric, we now show that the operator $\tilde{H}_{\Anti}$ is essentially self-adjoint, too, and its self-adjoint extension $H_{\Anti}$ is the restriction of $H$ to $\Anti L^2(\RRR^{Nd}) \cap\domain(H)$. Indeed, since $\tilde{H}$ commutes with every permutation, so does $H$; thus, both $\tilde{H}$ and $H$ commute with $\Anti$, which means they preserve anti-symmetry. It follows that
\be
H = \Anti H \Anti + (I-\Anti) H (I-\Anti)
\ee
and that $\Anti H \Anti$ and $(I-\Anti)H(I-\Anti)$ are self-adjoint operators. The operator $\Anti H\Anti$ defines a self-adjoint operator $H_{\Anti}$ on $\Anti L^2(\RRR^{Nd})$, which is a self-adjoint extension of $\tilde{H}_{\Anti}$. If $\tilde{H}_{\Anti}$ had any further self-adjoint extension $H'$ then 
\[
H' \Anti + (I-\Anti)H(I-\Anti)
\]
on the domain $\domain(H') \oplus \bigl(\domain(H)\cap (I-\Anti) L^2(\RRR^{Nd}) \bigr)$ would be a further self-adjoint extension of $\tilde{H}$, which we know does not exist.

Next we show that $U^{-1}H_{\Anti}U$ is a self-adjoint extension of $\tilde{H}_F$. Since the covering fibers $\proj^{-1}(q)$ are finite, the lift $\proj^{-1}(A)$ of any compact set $A$ is compact. As a consequence (using the remark after the proof of Proposition~\ref{prop:smooth}), $U$ maps $C_0^\infty(F)$ to $\Anti C_0^\infty(\cid)$, in fact bijectively. Note that $\Anti C_0^\infty(\cid)$ is a \emph{proper} subspace of $\Anti C_0^\infty(\RRR^{Nd})$, as a compact subset $A$ of $\RRR^{Nd}$ need not be a subset of $\cid$, and the intersection $A\cap \cid$ need not be compact; in other words, the domain of $U\tilde{H}_F U^{-1}$ is smaller than the domain of $\tilde{H}_{\Anti}$. Since $U$ can be regarded as the composition of first lifting a cross-section from $F$ to $\wh{F}$ and then using the bundle isomorphism $\hat{I}$ to transfer it to $\cid\times\CCC$, the Laplace operator on $F$ (defined on $C_0^\infty(F)$) gets mapped to the Laplace operator on the trivial Hermitian bundle $\cid\times\CCC$, which is the usual Laplacian on $\cid\subseteq \RRR^{Nd}$. Likewise, $U$ maps multiplication by $V$ to multiplication by $V$. Thus, $U$ maps $\tilde{H}_F$ to the restriction of $\tilde{H}_{\Anti}$ to $\Anti C_0^\infty(\cid)$, so that $H_{\Anti}$ is an extension of $U\tilde{H}_FU^{-1}$.
\end{myproof}

In some cases, $H_F$ is the \emph{only} self-adjoint extension of $\tilde{H}_F$. Obviously, this happens if and only if $-\tfrac{\hbar^2}{2m} \Laplace + V$ on $\Anti C_0^\infty(\RRR^{Nd}_{\neq})$ is essentially self-adjoint. This is the case in dimension $d\geq 4$, as the Laplace operator is essentially self-adjoint on $C_0^\infty(\RRR^p\setminus X)$ when $X\subset \RRR^p$ is a submanifold (or union of finitely many submanifolds) of codimension $\geq 4$ \cite{svendsen}. In any dimension $d$, the question is connected to the possibility of point interactions (i.e., range-zero interactions) such as
\be
V(\vx_1,\ldots,\vx_N) = \sum_{1\leq i<j\leq N} c\, \delta^d(\vx_i-\vx_j)\,,
\ee
where $\delta^d$ denotes the $d$-dimensional Dirac delta function and $c\in\RRR$ is a constant. Point interactions are impossible for $d\geq 4$ and possible in lower dimension; for $d=2,3$ they are possible only for bosons, not for fermions, in correspondence to the fact that the Laplacian in $\RRR^3$ away from the origin is not essentially self-adjoint only in the sector with angular momentum 0 \cite[pages 13 and 98]{AGHKH88}. We thus believe that $\tilde{H}_F$ is essentially self-adjoint for $d=2,3$. The situation changes when spin is introduced \cite{GTTZ14b}.

\setcounter{merken}{\theprop}
\setcounter{prop}{\thepropQUQU}
\begin{prop}
$U$ maps the position observable $\pvm_F$ to the position observable $\pvm_{\Anti}$,
$\pvm_F = U^{-1} \pvm_{\Anti} U$.
\end{prop}
\setcounter{prop}{\themerken}

\begin{myproof}
To begin with, $\pvm_{\Anti}(A)$ is indeed an operator on $\Anti L^2(\RRR^{Nd})$, i.e., $\pvm_{\Anti}(A)\wh{\psi}$ is indeed anti-symmetric if $\wh{\psi}$ is, since $1_{\proj^{-1}(A)}$ is a symmetric function, as the set $\proj^{-1}(A)$ is invariant under permutations. It is easy to see that $\pvm_{\Anti}(A)$ is a projection, that $\pvm_{\Anti}(\nrd)$ is the identity operator, and that $\pvm_{\Anti}$ is $\sigma$-additive; thus, $\pvm_{\Anti}$ is a PVM. So is $\pvm_F$. From \eqref{Uexplicit} we see that
\[
\pvm_{\Anti}(A)\, U\, \psi(\covq) =
1_{\proj^{-1}(A)}(\covq)  
\frac{1}{\sqrt{N!}}\hat{I}_{\covq} \psi(\proj\covq)=
\]
\be
=1_A(\proj\covq)  
\frac{1}{\sqrt{N!}}\hat{I}_{\covq} \psi(\proj\covq)=
U\, \pvm_F (A)\, \psi(\covq)\,,
\ee
which proves the desired equation.
\end{myproof}

\section{Quantum Triples}
\label{sec:equitriple}

In this section we argue for the statement \eqref{thesis}, which we repeat here for convenience: 
\setcounter{merken}{\theequation}
\setcounter{equation}{\thethesisnumber}
\begin{equation}
\mbox{
\begin{minipage}{0.85\textwidth}
\textit{In order to completely specify the mathematical data defining a quantum theory, it is necessary and sufficient to specify an equivalence class of quantum triples.}
\end{minipage}}
\end{equation}
\setcounter{equation}{\themerken}
To further elucidate the content of this statement, we should say that while an equivalence class of quantum triples provides all \emph{mathematical} data, it does not provide any data about the \emph{ontology}. Simply put, a quantum triple does \emph{not} select any particular interpretation of quantum theory, such as many-worlds \cite{Vai08}, Bohmian mechanics \cite{Gol01}, or a collapse theory (\`a la Ghirardi--Rimini--Weber--Pearle \cite{Ghi07}). The latter surely are different theories, and since they are theories of quantum mechanics it seems fair to call them different quantum theories. But they are not different examples of what is meant by ``quantum theory'' in \eqref{thesis}. They are alternative candidate theories for the same type of quantum system, they lead to (essentially) the same quantum formalism and the same predictions, and they propose different ontologies. In contrast, the different quantum theories in the sense of \eqref{thesis} apply to different types of systems, represent different versions of the quantum formalism, lead to different predictions, and say nothing about ontology. Examples of different quantum theories in this sense are provided by: bosons as different from fermions; spin-$\tfrac12$ particles as different from spinless particles; and different Hamiltonians, for example involving different interaction potentials.

Note that as a consequence of the thesis \eqref{thesis}, the fact that the fermionic line bundle is defined only up to isomorphy of Hermitian bundles is unproblematical, as isomorphic bundles give rise to the same quantum theory by \eqref{EE'equi}. In a different context, namely for providing a mathematical definition of the concepts of \emph{resonance} and \emph{resonant states}, ideas similar to the thesis \eqref{thesis} have been developed by Costin and Huang \cite{CH09}.

One thought behind \eqref{thesis} is that $\Hilbert$ and $H$ together, without the configuration observable $\pvm$, would not suffice for defining a quantum theory. After all, different choices of the position (or configuration) observable would lead to different quantum theories, making different predictions. If any example is necessary, consider the  hydrogen Hamiltonian $H= p^2 - \tfrac{1}{|q|}$ on $L^2(\RRR^3)$ with $p$ the usual 3 momentum and $q$ the 3 position operators; it is unitarily equivalent to $H'= q^2 - \tfrac{1}{|p|}$, which however has very different physical consequences; indeed, if we replaced $q$ by $\Fourier^{-1} q\Fourier$ with $\Fourier$ the Fourier transformation then in the new position representation $H$ would appear like $H'$.

Another thought behind \eqref{thesis} is that further observables need not be specified once the quantum triple is specified; that is, it is not an independent information
what the momentum operators are, the angular momentum operators etc. After all, at the end of \emph{any} experiment we read off results from \emph{positions} of things, such as positions of pointers or positions of ink droplets on paper. To appreciate this point, the reader should think of the quantum system as including any observers and apparatus. The probability distribution of configurations, of course, is fixed by a quantum triple together with an initial state vector $\psi_0\in\Hilbert$,
\begin{equation}\label{rhopsipvm}
  \rho_t(\cdot) = \scp{\psi_0}{e^{iHt/\hbar}\pvm(\cdot)e^{-iHt/\hbar} | \psi_0} \,,
\end{equation}
and indeed is the same for any equivalent quantum triple when $\psi_0$ is transformed accordingly. In particular, all empirical predictions can be derived once an equivalence class of quantum triples is specified. (And conversely, inequivalent quantum triples make
inequivalent predictions, in principle.)

\bigskip

Finally, we can actually prove \eqref{thesis} in our favorite formulation of quantum mechanics: Bohmian mechanics \cite{Bohm52,survey,Gol01}, a quantum theory without observers that solves the paradoxes and ambiguities of ordinary quantum mechanics by means of postulating an objective reality that exists independently of observers. In fact, this objective reality consists of particle trajectories, mathematically corresponding to a trajectory in the appropriate configuration space $\Q$, guided by the wave function $\psi_t$ according to a law of motion of the form
\begin{equation}\label{dQdtv}
  \frac{dq_t}{dt} = v(q_t) \,,
\end{equation}
where $q_t \in \Q$ is the actual configuration at time $t$ and $v=v[\Hilbert,H,\pvm,\psi_t]$ is a vector field on $\Q$, the velocity vector field. It can be expressed explicitly
in terms of the data $\Hilbert,H,\pvm,\psi_t$; namely, writing the vector field as a first-order differential operator acting on a test function $f\in C_0^\infty(\gencon)$:
\be\label{vdef}
v \cdot \nabla f(q) =   \mathrm{Re} \, \frac{\scp{\psi_t} {\pvm(dq) \frac{i}
   {\hbar} [H,\hat{f}] |\psi_t}} {\scp{\psi_t}{\pvm(dq)|\psi_t}}
\ee
with
\be
     \hat{f} = \int\limits_{q\in\Q} f(q) \, \pvm(dq) \,.
\ee
(To be sure, not every quantum triple
and state vector $\psi_t$ define a velocity vector field $v$, but when so
then they define $v$ uniquely.) The probability distribution of $q_t$ is given by the Born rule:
\be\label{Born}
\mathrm{Prob}(q_t \in A) = \scp{\psi_t}{\pvm(A)|\psi_t}\,.
\ee
Equations \eqref{dQdtv} and \eqref{vdef} are complemented by the usual Schr\"odinger time evolution of $\psi_t$,
\be\label{Schr}
\psi_t = e^{-iHt/\hbar}\psi_0\,.
\ee

\begin{prop}\label{prop:Bohm}
Suppose $(\Hilbert',H',\pvm')\sim(\Hilbert,H,\pvm)$, let $U:\Hilbert\to\Hilbert'$ be the unitary isomorphism that provides the equivalence, and set $\psi_t'=U\psi_t$. When replacing $\Hilbert$ by $\Hilbert'$, $H$ by $H'$, $\pvm$ by $\pvm'$ and $\psi_t$ by $\psi_t'$, then the Schr\"odinger evolution \eqref{Schr} is still valid, and the probability distribution \eqref{Born} and the right hand side of \eqref{vdef}, as an operator on $f\in C_0^\infty(\gencon)$, do not change. 
\end{prop}

Thus, whenever the vector field $v$ on $\gencon$ is well defined by \eqref{vdef} for all $t\in\RRR$ and whenever almost all solutions of the equation of motion \eqref{dQdtv} exist for all $t\in\RRR$ (see \cite{BDGPZ95,TT04} for conditions under which that is the case), any equivalent quantum triple leads to the same possible trajectories and the same probability distribution over the possible trajectories. We express this by saying that the Bohmian theories defined by $(\Hilbert',H',\pvm')$ and $(\Hilbert,H,\pvm)$ are \emph{physically equivalent}.

\bigskip

\begin{myproofof}{Proposition~\ref{prop:Bohm}}
Using \eqref{UHH'}, observe that 
\[
\psi'_t = U\,e^{-iHt/\hbar}\psi_0 = U\,e^{-iHt/\hbar} U^{-1}\psi'_0 = e^{-iH't/\hbar} \psi'_0\,.
\]
Furthermore, for any measurable $A\subseteq \gencon$,
\[
\scp{\psi'_t}{\pvm'(A)|\psi'_t}=\scp{U\psi_t}{U\pvm(A)U^{-1}|U\psi_t}=
\scp{\psi_t}{\pvm(A)|\psi_t}\,.
\]
Furthermore,
\[
\hat{f}' := \int\limits_{q\in\Q} f(q) \, \pvm'(dq)
=\int\limits_{q\in\Q} f(q) \, U\pvm(dq)U^{-1} = U\hat{f}U^{-1}
\]
and thus
\[
\mathrm{Re} \, \frac{\scp{\psi'_t} {\pvm'(dq) \frac{i}
   {\hbar} [H',\hat{f}'] |\psi'_t}} {\scp{\psi'_t}{\pvm'(dq)|\psi'_t}}=
\]
\[
=\mathrm{Re} \, \frac{\scp{U\psi_t} {U\pvm(dq)U^{-1} \frac{i}
   {\hbar} [UHU^{-1},U\hat{f}U^{-1}] |U\psi_t}} {\scp{U\psi_t}{U\pvm(dq)U^{-1}|U\psi_t}}=
\]
\[
=\mathrm{Re} \, \frac{\scp{\psi_t} {\pvm(dq) \frac{i}
   {\hbar} [H,\hat{f}] |\psi_t}} {\scp{\psi_t}{\pvm(dq)|\psi_t}}\,.
\]
\end{myproofof}

We close this section with a remark about the appropriate choice of $\Q$ in Bohmian mechanics. For $N$ identical particles in $\RRR^d$, $\Q=\nrd$ is the natural configuration space, also (and particularly) in Bohmian mechanics. For the sake of simplicity, one often uses $\RRR^{Nd}$ as the configuration space in Bohmian mechanics; this does not cause a problem for the following reason: As a consequence of the permutation symmetry or anti-symmetry of the wave function, the velocity vector field $v=(\boldsymbol{v}_1,\ldots,\boldsymbol{v}_N)$ in $\RRR^{Nd}$ is permutation-covariant,
\be
\boldsymbol{v}_{\sigma(k)}(\vx_{\sigma(1)},\ldots, \vx_{\sigma(N)}) 
= \boldsymbol{v}_k(\vx_1,\ldots, \vx_N)\,.
\ee
As a consequence of \emph{that}, if $\hat{q}_0,\hat{r}_0\in \RRR^{Nd}$ are two different orderings of the same unordered configuration $q_0\in\nrd$, i.e., if $\hat{q}_0 = \sigma \hat{r}_0$ for some permutation $\sigma \in S_N$, then the two solutions of the equation of motion \eqref{dQdtv} starting at $\hat{q}_0$ and $\hat{r}_0$ satisfy
\be
\hat{q}_t = \sigma \hat{r}_t
\ee
for all times $t\in \RRR$ with the same permutation $\sigma$. Thus, the arbitrary ordering of the initial configuration had no effect on the later configuration except on its ordering; put differently, the two trajectories $t\mapsto \hat{q}_t$ and $t\mapsto \hat{r}_t$ in $\RRR^{Nd}$ project to the same trajectory $t\mapsto q_t=\pi(\hat{q}_t) = \pi(\hat{r}_t)$ in $\nrd$, in fact to the trajectory we would have obtained from solving the equation of motion \eqref{dQdtv} directly in $\nrd$. In particular, the use of $\RRR^{Nd}$ as the configuration space has not affected the $N$ trajectories in $\RRR^d$. Likewise, by the permutation (anti-)symmetry of the wave function, the probability distribution it defines on $\RRR^{Nd}$ is permutation invariant and thus projects unambiguously to $\nrd$.

\section{Conclusions}
\label{sec:conclusions}

In this paper, we have provided a deeper justification and further development of the concept, introduced by Leinaas and Myrheim \cite{LM77}, of the fermionic line bundle. 
We have developed a general notion of equivalence of two quantum theories on a given configuration space; the use of this notion may go way beyond the discussion of identical particles and the symmetrization postulate. We propose that a quantum theory is specified by means of a ``quantum triple,'' i.e., a triple $(\Hilbert,H,Q)$ consisting of a Hilbert space $\Hilbert$, a Hamiltonian $H$, and a position PVM $Q$ on the configuration space acting on $\Hilbert$. According to our thesis, equivalent quantum triples describe the same quantum theory. We have provided a physical justification of this thesis as well as philosophical perspective.

We have proved that the quantum triple corresponding to cross-sections of the fermionic line bundle is equivalent to the one corresponding to the conventional description of fermionic wave functions, i.e., anti-symmetric functions on the space of ordered configurations. 

We believe that the space of ordered configurations is unphysical while the space of unordered configurations is natural. It may have seemed, however, that the unordered configurations cannot be used for fermions due to mathematical obstacles. Leinaas and Myrheim have shown how these obstacles can be overcome. Even more, it seems to us that for representing fermionic quantum states, cross-sections of the fermionic line bundle are more natural than anti-symmetric functions of ordered configurations.

\appendix

\section{Bundles With Given Holonomy Representation}
\label{app:representation}

We give here a proof of Proposition~\ref{prop:representation} and begin by repeating the statement.

\setcounter{merken}{\theprop}
\setcounter{prop}{\theproprepresentation}
\begin{prop}
For every connected differentiable manifold $\Q$, every $q\in\Q$, every $n\in\NNN$, and every $n$-dimensional unitary representation of the group $\pi_1(\Q,q)$, there exists a flat Hermitian bundle $E$ of rank $n$ over $\Q$ whose holonomy representation $\gamma_{E,q}$ is the given one. Any two such bundles are isomorphic; also, if two unitary representations of $\pi_1(\Q,q)$ are equivalent, then their associated flat Hermitian bundles are isomorphic. 
\end{prop}
\setcounter{prop}{\themerken}

\begin{myproof}
We first prove the second half of the second sentence; the first half of the second sentence then follows. More precisely, given
two flat Hermitian bundles $E,E'$ over $\Q$ whose holonomy representations $\gamma_{E,q}$, $\gamma_{E',q}$ are equivalent, we show that they are isomorphic. Choose a unitary isomorphism $I_q:E_q\to E'_q$ that provides an equivalence between $\gamma_{E,q}$ and $\gamma_{E',q}$, i.e., 
\be\label{gammaE'}
\gamma_{E',q}([\closedcurve]) = I_q \gamma_{E,q}([\closedcurve]) I_q^{-1}
\quad\text{for all }[\closedcurve] \in \pi_1(\Q,q)\,.
\ee
For every other point $r\in\Q$, choose a curve $\opencurve$ from $q$ to $r$ ($\Q$ is connected!) and define $I_r:E_r\to E'_r$ by parallel transporting $I_q$ along $\opencurve$. In fact, $I_r$ is independent of the choice of $\opencurve$, since parallel transport of $I_q$ along any closed curve $\closedcurve$ starting and ending at $q$ brings it back unaltered, as follows from $\holonomy'_\closedcurve I_q \holonomy_\closedcurve^{-1}=I_q$ or $\holonomy'_\closedcurve = I_q \holonomy_\closedcurve
I_q^{-1}$, which follows from \eqref{gammaE'}. Since $I$ is parallel, it is an isomorphism of \herm\ bundles.

We now turn to the first sentence. Let $W$ be an $n$-dimensional Hermitian vector space (i.e., Hilbert space) and $\gamma$ a unitary representation of $\pi_1(\Q,q)$ on $W$. The fiber $E_q$ of the bundle we will construct can be taken to be $W$, but it is clear that the substance of the proof is merely about constructing a bundle of the right isomorphy type, so this is what we will do.

We begin with a general consideration. If $M$ is a differentiable manifold and $G$ is a group acting properly  
discontinuously on $M$, then $M/G$ is a manifold and the projection $\proj:M\to M/G$ is a local diffeomorphism and, in fact, a covering map. Now consider a vector bundle $B$ over $M$
and an action of $G$ on $B$ such that each $g\in G$ maps, for any $x\in M$, the fiber $B_x$  
linearly to the fiber $B_{g(x)}$. Then the action of $G$ on $B$ is also properly  
discontinuous and $B/G$ is a manifold and, in fact, a vector bundle over $M/G$ in a natural way. To obtain the connection and inner products, note that any given $G$-invariant structure on $M$  
or on $B$ passes to the quotient and defines a corresponding structure  
on $M/G$ or $B/G$, respectively. So, for instance, a  
$G$-invariant connection on $B$ yields a connection on $B/G$ and a  
$G$-invariant inner product on $B$ yields an inner product on $B/G$. 

Now let us specialize to $M=\covspa$, the univeral covering space of $\Q$, and take $G$ to be the covering group, which consists of the deck transformations (i.e., the diffeomorphisms $g:\covspa\to \covspa$ that preserve fibers, i.e., $\pi(g(\covq))=\pi(\covq)$ for all $\covq\in\covspa$). It is known that the covering group $G$ is canonically isomorphic to the fundamental group $\pi_1(\Q,q)$ for any choice of $q$ (viz., the isomorphism $\iota_q$ maps $[\closedcurve]$ to the unique deck transformation $g=\iota_q([\closedcurve])$ such that $g(\covq)=\covq'$ for any lift $\hat{\closedcurve}$ of $\closedcurve$ to $\covspa$ with starting point $\covq$ and end point $\covq'$). Furthermore, $G$ acts properly discontinuously on $\covspa$, $M/G=\Q$, and the projection $\proj$ coincides with the natural covering map $\covspa\to\Q$. Take $B$ to be the trivial Hermitian bundle $B=\covspa\times W$ and the action of $G$ on $B$ to be $\gamma(\iota_q^{-1}(g)):B_x\to B_{g(x)}$ for every $g\in G$ and every $x\in\covspa$ (recall $B_x=W=B_{g(x)}$). Then the inner product and connection are $G$-invariant, and $E=B/G$ is the desired Hermitian bundle. Indeed, to compute $h_\closedcurve$ in $E$ for a closed curve $\closedcurve$ based at $q$, let $\hat{\closedcurve}$ be a lift of $\closedcurve$ starting at $\covq_0\in\proj^{-1}(q)\subset\covspa$ and ending at $\covq=g (\covq_0)$. Parallel transport of an element $\phi\in B_{\covq_0}$ along $\hat{\closedcurve}$ leads to the same element of $B_{\covq}=W=B_{\covq_0}$ because the connection of $B$ is trivial, and since $B_{\covq}$ and $B_{\covq_0}$ are identified with $E_q$ in different ways, differing by $\gamma([\closedcurve])$, the holonomy is $h_\closedcurve = \gamma([\closedcurve])$, which completes the proof.
\end{myproof}

\bigskip

\noindent\textit{Acknowledgments. }
We thank Detlef D\"urr (M\"unchen), Frank Loose (T\"ubingen), Daniel Victor Tausk (S\~ao Paulo), and Stefan Teufel (T\"ubingen) for helpful discussions. This research was supported by mini-grant \#MGA-09-013 from The Foundational Questions Institute (fqxi.org). N.Z.\ gratefully acknowledges support by INFN and the European Cooperation in Science and Technology (COST action MP1006). 
Finally, we appreciate the hospitality that some of us have enjoyed,
on more than one occasion, at the Mathematisches Institut of
Ludwig-Maximilians-Universit\"at M\"unchen (Germany), the Dipartimento
di Fisica of Universit\`a di Genova (Italy), the Institut des Hautes
\'Etudes Scientifiques in Bures-sur-Yvette (France), and the
Mathematics Department of Rutgers University (USA).

\end{document}